\documentclass[preprint,1p]{elsarticle}

\usepackage{amsmath,amssymb,mathtools}
\usepackage{hyperref}
\usepackage{setspace}
\usepackage{lineno}
\usepackage{xcolor}
\usepackage{booktabs}
\usepackage{multirow}
\usepackage{calc,array}
\usepackage{longtable}
\usepackage{supertabular}
\usepackage{adjustbox}
\usepackage{colortbl}
\usepackage{float}

\journal{Pattern Recognition}

\providecommand{\tightlist}{%
  \setlength{\itemsep}{0pt}\setlength{\parskip}{0pt}}

\newenvironment{widetable}[1][]%
{%
    \renewcommand{\endhead}{}
    \renewenvironment{longtable}{\begin{center}\caption{#1}\begin{supertabular}}{\end{supertabular}\end{center}}
    \let\columnwidth\textwidth
    \begin{table*}[!ht]
    \scriptsize
}%
{%
    \end{table*}
}

\newcommand{\hideFromPandoc}[1]{#1}

\hideFromPandoc{%

}

{%
    \makeatletter
    \def\maxwidth{\ifdim\Gin@nat@width>\linewidth\linewidth\else\Gin@nat@width\fi}
    \def\maxheight{\ifdim\Gin@nat@height>\textheight\textheight\else\Gin@nat@height\fi}
    \makeatother
    \setkeys{Gin}{width=\maxwidth,height=\maxheight,keepaspectratio}
    \renewenvironment{figure}{\begin{figure*}[t]\centering}{\end{figure*}}
}{}

\newcommand\pmi{\ensuremath{\text{PMI}}}

\newcommand\logq{\ensuremath{\log Q}}
\newcommand\sigmoidq{\ensuremath{\sigma(Q)}}

\newcommand\truncatelogq{\ensuremath{\log[\max(Q, 1)]}}
\newcommand\words{\ensuremath{w_1, w_2, \ldots, w_n}}

\newcommand\rocaucscore{\ensuremath{\text{ROC AUC score}}}

\begin{document}

\begin{frontmatter}

\title{Neural graph embeddings as explicit low-rank matrix factorization
for link prediction}



\author[1]{Asan Agibetov \corref{4}}
\cortext[4]{Corresponding author; Währinger Straße 25A, OG1.06, 1090,
Vienna, Austria}
\ead{asan.agibetov@meduniwien.ac.at}

\address[1]{Institute of Artificial Intelligence, Medical University of
Vienna, Vienna, Austria}

\begin{abstract}%
\noindent Learning good quality neural graph embeddings has long been
achieved by minimzing the pointwise mutual information (PMI) for
co-occuring nodes in simulated random walks. This design choice has been
mostly popularized by the direct application of the highly-successful
word embedding algorithm word2vec to predicting the formation of new
links in social, co-citation, and biological networks. However, such a
skeuomorphic design of graph embedding methods entails a truncation of
information coming from pairs of nodes with low PMI. To circumvent this
issue, we propose an improved approach to learning low-rank
factorization embeddings that incorporate information from such unlikely
pairs of nodes and show that it can improve the link prediction
performance of baseline methods from 1.2\% to 24.2\%. Based on our
results and observations, we outline further steps that could improve
the design of next graph embedding algorithms that are based on matrix
factorizaion.
\end{abstract}


\end{frontmatter}


\hypertarget{introduction}{%
\section{Introduction}\label{introduction}}

Networks are a convenient representation of the relationships of complex
system components, such that the interconnectivity of nodes and edges
inherently encodes valuable information about the system. It is such a
flexible tool that has been used to represent complex emerging patterns
in disparate domains, from molecular biology to social sciences
\citep{reka2002}. Perhaps, one reason for the flexibility of this
representation, is that the information stored within can be understood
through the analysis of its topology. \emph{Link
prediction}~\citep{libennowell2003} has become a popular pattern
recognition problem, where given the current topology of the network we
predict the next most likely links to emerge. To address this problem,
one common way is to treat the network as a graph \(G=(V, E)\) with
nodes \(u \in V\) and edges \((u, v) \in E; u, v \in V\). Then, to
perform link prediction, one would split the graph edges \(E\) into two
partitions (not necessarily of the same size) \(E_{train}, E_{test}\),
which serve the purpose of \emph{positive} edges. Next, we sample the
non-existent edges \((u, v) \not \in E\) to make
\(\bar{E_{train}}, \bar{E_{test}}\), which serve as \emph{negative}
examples. Habitually, one would then come up with a prediction model
that would learn patterns from the training dataset; typically learn to
score positive edges higher than negatives. And the prediction
performance of such a model would be assessed on the test set with an
accuracy metric. As a result, one would hope that through this
partitioning the prediction model would generalize enough to predict the
future states of the system, i.e., to predict the formation of new links
\citep{ma2022}.

The conventional way to efficiently process the network structure to
analyze the complex system relies on explicit topological descriptors,
such as betweenness centrality and triangle count. The accuracy of such
ad-hoc descriptors is directly proportionate to the expertise and
domain-knowledge put into their design. However, recently, the field of
representation learning has proposed a more flexible approach of
learning latent representation of networks, which embeds the graph
structure into a latent space. More specifically, given a graph \(G\)
the goal is to learn a dictionary mapping \(V \mapsto \mathbb{R}^d\)
that embeds each vertex to a \(d\)-dimensional vector. The link
prediction is then performed by a classifier
(\(V \times V \mapsto \mathbb{R})\) that for a pair of input vertices
\(u, v\) scores the likelihood of them being connected (i.e., likelihood
to form a link).

The roots of this line of work can be traced back to the spectral graph
theory~\citep{chung1997} and social dimensional
learning~\citep{tang2009}. However, the more recent approaches (e.g.,
\emph{deepwalk}~\citep{perozzi2014}, \emph{node2vec}~\citep{grover2016})
have been based on word embedding algorithms, in particular the
skip-gram with negative sampling (SGNS) version of the \emph{word2vec}
model~\citep{mikolov2013word2vec}. The central idea is that the node
sequences \(\words\) generated from random walks simulation on graphs
can be treated as a text corpus and inputted into the word2vec model to
get the output in a form of latent vector representation for each node
(technically a word in the random walks generated text). While
deepwalk-inspired approaches have been demonstrated to be effective in
link prediction problems, their theoretical mechanisms have been
relatively understudied, up until Qiu et al.~\citep{qiu2018} have
explicitly demonstrated the connection of deepwalk-inspired approaches
and matrix factorization methods. The central idea relies on the fact
that the horseback of these approaches -- word2vec -- has been shown to
implicitly factorize the matrix containing the collocation of words in
the text. Concretely, Levy et al~\citep{levy2014} have formally
demonstrated that the word2vec is factorizing the point-wise mutual
information (PMI) matrix - a well-known matrix in the information
theory. Based on this demonstration, Qiu et al.~\citep{qiu2018} have
given closed-form expressions for the PMI matrix that deepwalk-inspired
methods are factorizing in graph-theoretic terms. This expression of the
PMI matrix is dubbed the \emph{deepwalk matrix} and its closed-form
derivation is tightly connected to random walks and graph Laplacians.
Finally, the authors proposed the netmf algorithm that learns network
embeddings for nodes by performing the singular value decomposition
(SVD) of the deepwalk matrix.

\hypertarget{challenges}{%
\subsection{Challenges}\label{challenges}}

An important observation in this general approach to low-rank
approximation of the PMI matrix is that this matrix may pose
computational issues. The trouble kicks in when we have a pair of words
\((w, c\)) which are never observed. These \emph{negative} pairs lead to
the ill-posed task of decomposing \(\pmi (w, c) = \log 0 = - \infty\)
entries. As noted in~\citep{levy2014}, solutions to these problems
include a Dirichlet prior by adding a small ``fake'' count to the
underlying counts matrix - smoothing thus negative entries, or
truncating the PMI matrix, such that \(\pmi (w, c) = 0\) for the
unobserved pairs. Qiu et al.~\citep{qiu2018} use the latter, mainly
because they want to leverage the fact that a truncated matrix is very
sparse, and thus SVD is accelerated. However, by forcing zeroes into the
matrix we ignore the information from most of the entries of the matrix,
as PMI scores for low-frequency pairs of nodes will be indistinguishable
from pairs of nodes that truly never appear in contexts. We show that
this may lead to a drop in performance on link prediction problems.

\hypertarget{contribution-of-this-work}{%
\subsection{Contribution of this work}\label{contribution-of-this-work}}

In this work, we extend the work of Qiu et al.~\citep{qiu2018} and
propose a new approach to learning low-rank factorization embeddings
that incorporate information from unlikely pairs of nodes. The
contributions of our work may be summarized in these three bullet
points:

\begin{itemize}
\tightlist
\item
  We apply smoothing to shifted PMI scores to enable a low-rank
  factorization of matrix entries that does not penalize low-frequency
  node pairs. We show that this correction may drastically improve
  accuracy.
\item
  By considering different transformations of the \(\pmi\) matrix, and
  by deriving matrix forms of the joint probability matrix, we
  demonstrate superior link prediction performance. These results lead
  us to argue that the linguistic collocation metric (PMI) may not be
  the best metric for link prediction.
\item
  Finally, we compare our results with the state-of-the-art accuracy
  scores from the approach of Abu-El-Haija et al~\citep{abu2018}, and
  conclude that our approach is on par. Their algorithm is using a
  stochastic (implicit) matrix factorization using neural networks,
  while ours is based on explicit SVD.
\end{itemize}

\hypertarget{organization-of-this-manuscript}{%
\subsection{Organization of this
manuscript}\label{organization-of-this-manuscript}}

This manuscript is organized as follows: in Section~\ref{sec:theory} we
will highlight the theoretical notions that are necessary for a fluid
reading of our work; Section \ref{sec:proposed-approach} will present
our approach; in Section~\ref{sec:methods} we will review the specific
steps we took to evaluate our approach on link prediction tasks;
Section~\ref{sec:results} will cover our results; finally, we will close
with a discussion section (Section~\ref{sec:discussion}).

\hypertarget{sec:theory}{%
\section{Theory}\label{sec:theory}}

To improve the self-contained reading of this manuscript, we briefly
highlight the most salient parts of the required theoretical concepts.

\hypertarget{neural-word-embeddings-with-skip-gram-negative-sampling}{%
\subsection{Neural word embeddings with skip-gram negative
sampling}\label{neural-word-embeddings-with-skip-gram-negative-sampling}}

Neural graph embedding research started with the introduction of
\emph{deepwalk}~\citep{perozzi2014} algorithm, which adapted the famous
word representation algorithm \emph{word2vec}
\citep{mikolov2013word2vec} to latent graph representation learning.
Recall that, the goal of \emph{word2vec} is to learn a dictionary
mapping \(\Theta: W \mapsto \mathbb{R}^d\), which embeds a word
\(w_i \in W\) into a \(d\)-dimensional vector. These vectors encode
generic linguistic information of words in a text corpus and can be used
in downstream tasks, such as document classification.

Given a word sequence \(w_{c_1}, w_{c_2}, w_t, w_{c_3}, w_{c_4}\), a
word2vec version based on \emph{skip-gram} training objective tries to
predict \emph{context} words \(w_{c_i}\) given the \emph{target} word
\(w_t\). Formally, it models the conditional probability
\(P_{\Theta}(w_{c_i} | w_t) = \frac{e^{\langle \Theta(w_t), \Theta(w_{c_i}) \rangle}}{\sum_k e^{\langle \Theta(w_t), \Theta(w_k) \rangle}}\)
using inner products, e.g., Euclidean dot product, of word embeddings,
e.g.~\(\Theta(w_t)\), and a softmax function in the denominator. The
exact computation of the softmax function is computationally infeasible
for any real-world text corpus, i.e., \(k\) would range over millions of
words. To approximate the softmax function, for one target-context pair
of words \((w_t, w_c)\), Mikolov et al.~\citep{mikolov2013sgns} proposed
to sub-sample \(k\) words from a unigram context distribution
\(w_k \sim U\), which do not appear in the context of \(w_t\), and
minimize a negative log-likelihood loss function

\[
L_{(w_t, w_c)} = - \left[ \log \sigma(\langle \Theta(w_t), \Theta(w_c) \rangle) + \sum_{j \sim U}^k \log \left[ 1 - \sigma(\langle \Theta(w_t), \Theta(w_j) \rangle) \right] \right] \qquad (1),
\]

where a sigmoid function \(\sigma(x) = \frac{1}{1 + e^{-x}}\) squashes
everything into a \(0, \ldots, 1\) range.

This approximation is called skip-gram with negative sampling (SGNS). In
SGNS, model parameters \(\Theta\), e.g.~word embeddings, are learned
implicitly by training a neural network that minimizes an aggregated
negative log-likelihood,
\(\text{argmin}_{\Theta} -\sum_{(t, c) \in \Omega} L_{(w_t, w_c)}\), of
all target-context pairs \(\Omega\) in the training corpus. In this
model, words that appear in close proximity in the text would tend to
get embedded closer to each other in the embedding space.

\hypertarget{connection-to-neural-graph-embeddings}{%
\subsection{Connection to neural graph
embeddings}\label{connection-to-neural-graph-embeddings}}

The natural connection between graph and word embeddings relies in the
fact, that, given a graph \(G = \langle V, E \rangle\), and a starting
node \(w_t \in V\), we could launch a random walk of length \(l\) to
generate a context node sequence \(w_t, w_{c_1}, \ldots, w_{c_l}\).
Performing such random walk simulations for all nodes in the graph would
generate many context node sequences, which we could group in a multiset
\(\Omega\) consisting of target-context node pairs
\((w_t, w_c), w_t, w_c \in V\). Effectively, we could treat \(\Omega\)
as a training text corpus, and use SGNS word2vec model to learn a
dictionary mapping \(\Theta: V \mapsto \mathbb{R}^d\), for each node
\(w_i \in V\). Nodes that appear more frequently in the same context
node sequence would get embedded closer to each other. This is the
central idea of neural graph embedding models, that rely on word2vec,
such as deepwalk \citep{perozzi2014} and node2vec \citep{grover2016}.

\hypertarget{connection-to-information-theory-and-low-rank-matrix-approximation}{%
\subsection{Connection to information theory and low-rank matrix
approximation}\label{connection-to-information-theory-and-low-rank-matrix-approximation}}

Levy et al.~\citep{levy2014} have formally demonstrated that for a
target-context pair \((w_t, w_c)\), Eq. 1 will be minimized when the
inner product is equal to
\(\langle \Theta(w_t), \Theta(w_c) \rangle = \log \frac{P(w_t, w_c)}{P(w_t) P(w_c)} - \log b\),
where \(b\) is the amount of negative words for the target word \(w_t\).
This can be obtained by setting the gradient of with respect to the
inner product to zero,
\(\frac{\partial L(w_t, w_c)}{\partial \langle \Theta(w_t), \Theta(w_c) \rangle} = 0\),
and simplifying algebraically. In fact,
\(\text{PMI}(w_t, w_c) = \log \frac{P(w_t, w_c)}{P(w_t) P(w_c)}\) is
point-wise mutual information (PMI), a well-known quantity in the
information theory, which measures the discrepancy between the
probability of the coincidence of two random variables given their joint
distribution and their individual distributions. PMI is used in
linguistics to measure the collocation or linguistic association of two
words. Good collocation pairs have high PMI because the probability of
co-occurrence is only slightly lower than the probabilities of
occurrence of each word.

This theoretical result allows us to interpret the condition for the
optimized SGNS word2vec model. Effectively, the best model is obtained,
when for each target-context pair \((w_t, w_c)\) their inner product
approaches the shifted pointwise mutual information statistic

\[
\frac{\partial L(w_t, w_c)}{\partial \langle \Theta(w_t), \Theta(w_c) \rangle} = 0, \text{ when } \langle \Theta(w_t), \Theta(w_c) \rangle \to \text{PMI}(w_t, w_c) - \log b. \qquad (2)
\]

And, by grouping target-context pairs \((w_t, w_c)\) as entries in a
matrix \(M_{w_t, w_c} = \text{PMI}(w_t, w_c) - \log b\), we can treat
the minimization of the loss function in Eq. 1 as an implicit matrix
factorization of a huge matrix \(M\) filled with PMI statistics. We can
formalize it with

\[
\text{argmin}_{Y, R} \lVert M - Y \times R^T \rVert_F, \qquad (3)
\]

where \(Y, R\) are weight matrices of the hidden layer in the word2vec
neural network, and \(\lVert \cdot \rVert\) is the Frobenius norm. By
optimizing the weights \(Y, R\) of the neural network we implicitly find
word embeddings, such that their pairwise inner products approach the
shifted PMI (see Eq. 2).

\hypertarget{connection-to-random-walks-on-graphs}{%
\subsubsection{Connection to random walks on
graphs}\label{connection-to-random-walks-on-graphs}}

Since neural graph embeddings use SGNS word2vec to learn vector
representations for nodes, it was a natural step to re-interpret Eq. 2
in terms of intrinsic properties of a graph \(G\). Qiu et
al.~\citep{qiu2018} have re-expressed the shifted-PMI of a
target-context pair of nodes \((w_t, w_c)\) in graph-theoretic terms as,

\[
\text{PMI}(w_t, w_c) - \log b \to_p \log \left[ \frac{\text{vol}(G)}{2T} \left( \frac{1}{d_{w_c}} \sum_{r=1}^T (P^r)_{w_t, w_c} + \frac{1}{d_{w_t}} (P^r)_{w_c, w_t} \right) \right] - \log b, \qquad (4)
\]

where \(d_{w_t}\) is the degree of the node \(w_t\),
\(\text{vol}(G) = \sum_i d(i)\) is the volume of graph \(G\) (sum of all
degrees), \(T\) size of the context window of a random walk (how far to
the left and right we will search for context nodes), and \(b\) is the
number of negative nodes for a target node \(w_t\). The transition
probability matrix \(P = D^{-1} A\) is obtained from the adjacency \(A\)
and degree \(D\) matrices of a graph, such that an entry
\(P_{w_t, w_c}\) holds a probability of transitioning into \(w_c\) from
\(w_t\) in 1 step. Power matrices \(P^r = P(P( \ldots (P) \ldots)\) give
us conditional probabilities of transitioning from one node to others in
\(r\) steps, i.e., \(P^r_{w_t, w_c} = P(x_r = w_c| x_1 = w_t)\) for a
random walk \(x_1 = w_t, \ldots, x_r = w_c\).

Effectively, Qiu et al.~have shown that deepwalk is optimized when the
inner product of a target-context node pair \((w_t, w_c)\) is
approaching \(\text{PMI}(w_t, w_c) - \log b\). After a series of
derivations, they finally propose the closed form of the shifted PMI
matrix in terms of graph theory terminologies

\[
\begin{aligned}
M = \pmi - \log b &= \log \left[ \frac{\text{vol}(G)}{2T} \left( \frac{1}{d_{w_c}} \sum_{r=1}^T (P^r)_{w_t, w_c} + \frac{1}{d_{w_t}} (P^r)_{w_c, w_t} \right) \right] - \log b \\
                  &= \log \left( \frac{\text{vol}(G)}{T} \left( \sum_{r=1}^T (P^r) \right) D^{-1} \right) - \log b \\
              &= \log \left( \frac{\text{vol}(G)}{bT} \left( \sum_{r=1}^T (P^r) \right) D^{-1} \right) \qquad \text{rename argument of } \log \text{ as } Q \\
                  &= \log Q. \qquad (5)
\end{aligned}
\]

Throughout this manuscript, we will refer to \(\log Q\) as the deepwalk
matrix. As mentioned previously, the matrix \(M\) is ill-imposed, since
pairs of nodes that never occur in \(\Omega\) lead to
\(\log 0 = - \infty\). To circumvent this computational trouble, Qiu et
al~\citep{qiu2018} consider the truncated version
\(M' = \log (\max(Q, 1))\) leading to a highly sparse representation.
Consequently, they only consider pairs of nodes that occur in \(\Omega\)
during the decomposition process, or even more dramatically, they ignore
the information from most of the entries of \(M\). To obtain a
\(d\)-rank approximation of \(M \approx L \times R^T\), Qiu et
al.~\citep{qiu2018} apply singular value decomposition (SVD) directly on
\(M\), instead of implicitly learning them by minimizing Eq. 3 with a
neural network. By explicitly factorizing the closed-form of the
deepwalk matrix with SVD via \(M' = U_d \Sigma_d V_d^T\), NetMF finally
returns the matrix of node embeddings

\[
Y = U_d \sqrt{\Sigma_d}. \qquad(6)
\]

\hypertarget{sec:proposed-approach}{%
\section{Proposed approach}\label{sec:proposed-approach}}

The skip-gram powered neural graph embedding approaches relied on
word2vec to learn latent vector representation by implicitly factorizing
the shifted PMI matrix. Driven by the goal of theoretically explaining
deepwalk-like neural graph embedding algorithms, Qiu et
al.~\citep{qiu2018} settled for the PMI matrix, and especially its
closed-form expression in graph-theoretic terms. However, the choice of
the PMI matrix for learning graph embeddings was implicitly imposed by
the design of word2vec. In developing our approach, we ponder whether
the matrix factorization of PMI matrix is the best way to obtain node
representations suitable for link prediction of real-world networks? To
test our hypothesis, we first show why PMI matrix might be a bad option,
and then use this rationale to derive our method.

\hypertarget{contribution-of-low-frequency-pairs-to-learning-good-graph-embeddings}{%
\subsection{Contribution of low-frequency pairs to learning good graph
embeddings}\label{contribution-of-low-frequency-pairs-to-learning-good-graph-embeddings}}

\begin{figure}[!h]
    \begin{center}
        \includegraphics[width=1.\linewidth]{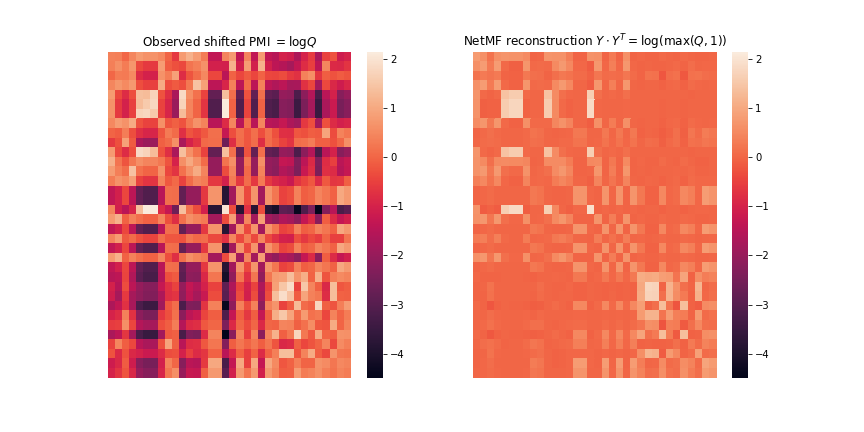}
    \end{center}
    \caption{\label{fig:netmf-truncatedPMI}Left: ground truth shifted PMI matrix for the Karate graph; 
             Right: NetMF ($\truncatelogq$) approximation. Low-rank factorization was computed with dim 5, neg size 1, window size 5. 
             Changing hyperparameter values does not change the fact that NetMF ($\truncatelogq$) ignores low frequency pairs.}
\end{figure}

We start our incursion into the effect of truncating or smoothing
negatives on link prediction by visually investigating the
reconstruction of the PMI matrix. First, we apply the original NetMF
(\(\truncatelogq\)) algorithm on the karate club network
\citep{karate1977}, consisting of 34 nodes, to obtain graph embeddings
\(Y\) (Eq. 6). Then, we compare the ground truth PMI matrix (Figure
\ref{fig:netmf-truncatedPMI}, left) and the reconstruction with
\(Y \cdot Y^T\) (Figure~\ref{fig:netmf-truncatedPMI}, right). We can see
that \(Y \cdot Y^T\) ignores the contribution of low-frequency pairs
\((u, v\)), where \(Q_{uv} < 1\). While NetMF (\(\truncatelogq\))
reconstruction is fairly good for highly frequent pairs of nodes, there
is little variation between the non-occurring pairs and the pairs that
occur with a low PMI.

\hypertarget{matrix-form-for-random-walk-context-window-co-occurrence-probability}{%
\subsection{Matrix form for random walk context window co-occurrence
probability}\label{matrix-form-for-random-walk-context-window-co-occurrence-probability}}

To test our hypothesis, we propose two ways to include the contribution
of truncated PMI entries. The first is a straightforward transformation
of the original deepwalk-matrix, where instead of truncating we are
smoothing the negatives by applying \(\sigma(\cdot)\) to the entries of
\(Q\) (Eq. 5). In doing so, we lose the semantics of the shifted
pointwise mutual information metric, because we are skipping the
\(\log\) transformation, however, we do include the contribution of
low-occurrence pairs. Our second approach is to consider the joint
probability of two nodes to appear in the same random walk context. For
this, we need to derive the matrix form of the joint distribution, as
shown below

\[
\begin{aligned}
J=&\frac{1}{2T} \sum_{r=1}^T \left( \frac{d_w}{\text{vol}(G)} (P^r)_{w, c} + \frac{d_c}{\text{vol}(G)} (P^r)_{c, w} \right) \\
& \qquad = \frac{1}{2T \cdot \text{vol}(G)} \left( \sum_{r=1}^T D (P^r) + \sum_{r=1}^T (P^r)^T D \right) \\
& \qquad = \frac{1}{2T \cdot \text{vol}(G)} \left( \sum_{r=1}^T \underbrace{D D^{-1}}_{I} \underbrace{A \times D^{-1}A \times \ldots \times D^{-1}}_{r-1 \text{ times}} A
\quad + \right. \\
& \qquad \qquad \qquad \qquad \left. + \quad \sum_{r=1}^T \underbrace{A D^{-1} \times \ldots A D^{-1}}_{r-1 \text{ times}} \times A \underbrace{D^{-1} D}_{I} \right) \\
& \qquad = \frac{1}{T \cdot \text{vol}(G)} \sum_{r=1}^T \underbrace{A D^{-1} \times \ldots A D^{-1}}_{r-1 \text{ times}} \times A \\
& \qquad = \frac{1}{T \cdot \text{vol}(G)} \left( \sum_{r=1}^{T-1} (P^r)^T \right) A. \qquad (7)
\end{aligned}
\]

This matrix form of \(J\) gives us a closed-form solution for random
walk co-occurrence probability computation, without the need for
generating random walk sentences and sampling context pairs.
Table~\ref{tab:matrices} gives an overview of the considered matrices,
derived from the original NetMF algorithm, and details what one entry in
each matrix means. Entries in these matrices represent statistic
quantity that can be computed from the original graph structure.

\begin{widetable}[\label{tab:matrices} Matrices whose SVD low-rank factorization is considered in our link prediction experiments. An entry $M_{ij}$ in a considered matrix $M$ records a statistic of co-occurrence of two nodes $i, j$; graph statistics are derived from random walks.]

\begin{longtable}[]{@{}lc@{}}
\toprule
matrix \(M\) & meaning of entry \(M_{ij}\)\tabularnewline
\midrule
\endhead
NetMF(\(\log[\max(Q, 1))\)) & truncated shifted point-wise mutual
information \(\log[\max(Q_{i, j}, 1))\) for nodes
\(i, j\)\tabularnewline
NetMF(\(\sigma(Q)\)) & \(\sigma(x) = \frac{1}{1 + e^{-x}}\) applied to
each entry of \(Q\) (smoothened negatives)\tabularnewline
\(J\) & \(P(i, j)\) - probability \(p(i, j)\) that nodes \(i, j\)
co-occur within the same context in all simulated random
walks\tabularnewline
\bottomrule
\end{longtable}

\end{widetable}

\hypertarget{sec:methods}{%
\section{Methods}\label{sec:methods}}

\hypertarget{datasets}{%
\subsection{Datasets}\label{datasets}}

To perform unbiased comparison of link prediction performance using
different graph embedding models, we used exactly the same train and
test splits as the ones used in Abu-El-Haija et al.~\citep{abu2018}. In
\citep{abu2018}, 5 networks from \footnote{\url{https://snap.stanford.edu/data/}}SNAP
(Stanford Large Network Dataset Collection) were considered, which we
summarize in Table \ref{tab:datasets}.

\begin{widetable}[\label{tab:datasets} Datasets used in all experiments. Note that the PPI dataset is no longer available from the SNAP website (Last checked: May 27, 2020). To use the same dataset that was tested in this work you would have to download the version available in~\cite{abu2018}. PPI dataset was originally published in~\cite{biogrid2006}.]

\begin{longtable}[]{@{}lcccc@{}}
\toprule
Dataset & Type & \(|V|\) & \(|E|\) & Description\tabularnewline
\midrule
\endhead
PPI~\citep{biogrid2006} & undirected & 3,852 & 38,705 & Protein-protein
interaction network\tabularnewline
ca-HepTh & undirected & 8,638 & 24,826 & Collaboration network of Arxiv
High Energy Physics\tabularnewline
soc-Facebook & undirected & 4,039 & 88,234 & Social circles from
Facebook (anonymized)\tabularnewline
wiki-Vote & directed & 7,066 & 103,663 & Wikipedia who-votes-on-whom
network\tabularnewline
ca-AstroPh & undirected & 17,903 & 197,031 & Collaboration network of
Arxiv Astro Physics\tabularnewline
\bottomrule
\end{longtable}

\end{widetable}

\hypertarget{graph-embedding-models}{%
\subsection{Graph embedding models}\label{graph-embedding-models}}

We compared the link prediction performance of our approach to the
original NetMF algorithm and to the following state-of-the-art graph
embedding models: BigClam \citep{yang2013bigclam}, NNSED
\citep{sun2014nmfadmm}, SocioDim \citep{tang2009sociodim}, RandNE
\citep{zhang2018randne}, NFMADMM \citep{sun2014nmfadmm}, GLEE
\citep{torres2020}, and NodeSketch \citep{yang2019nodesketch}. All of
our experiments were primarily implemented in Python (version 3.8). We
used the usual packages from the scientific Python stack, such as numpy,
scipy, and scikit-learn. To compare our approach with \textsc{NetMF}, we
used the \footnote{\url{https://github.com/xptree/NetMF}}official
implementation available on GitHub. For other models, we used karateclub
\citep{karateclub2020}, a Python library for unsupervized learning on
graph-structured data.

Furthermore, we used link prediction scores in \citep{abu2018} to
directly compare neural network-based graph embedding models to the
proposed method. Specifically, the link prediction performance of the
following models was considered: DGNR \citep{cao2016}, Node2Vec
\citep{grover2016}, Assym Proj \citep{abu2017}, and WatchYourStep
\citep{abu2018}. Note that we have not performed hyperparameter
optimization for these models, but we took the best scores from
\citep{abu2018}.

\hypertarget{link-prediction-evaluation-and-bayesian-hyperparameter-optimization}{%
\subsection{Link prediction, evaluation, and Bayesian hyperparameter
optimization}\label{link-prediction-evaluation-and-bayesian-hyperparameter-optimization}}

\begin{figure}[!h]
    \begin{center}
        \includegraphics[width=1.\linewidth]{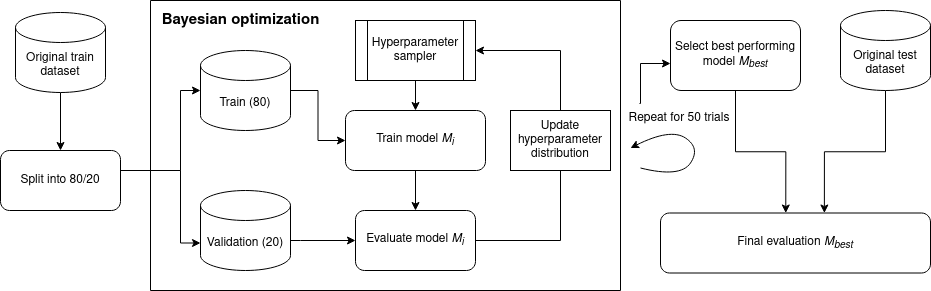}
    \end{center}
    \caption{\label{fig:bayesian-optimization}Bayesian optimization overview.}
\end{figure}

To make sure that all graph embedding models were compared head-to-head
in a fair manner, we performed the model selection, where we optimized
for the best hyperparameters for each model on every considered graph.
Since all considered graph embeddings models are sensitive to the
specific choice of the associated hyperparameters, model selection was
performed using a computationally intensive Bayesian hyperparameter
optimization pipeline summarized in Figure
\ref{fig:bayesian-optimization}. For example, our proposed approach has
only one parameter to optimize the deep walk length, i.e., the highest
power of the transition matrix. NetMF adds another parameter, the number
of negative samples in the computation of the pointwise mutual
information metric. Graph embedding models based on non-negative matrix
factorization have other hyperparameters, such as the number of
iterations to solve the NMF optimization problem. We provide the full
list of sampled hyperparameter configurations and the underlying
distributions for each graph embedding approach in Supplementary
Material. Our hyperparameter optimization pipeline was implemented using
\footnote{\url{https://github.com/optuna/optuna}}\texttt{optuna}, a
Python hyperparameter optimization library.

We took train and test splits provided in Abu-El-Haija et
al.~\cite{abu2018}. The original train set was used to learn graph
embeddings \(Y\) and perform hyperparameter tuning, and the original
test set was only used for the final link prediction evaluation of the
best selected model. For classifying positive and negatives links using
graph embeddings \(Y\), obtained with a graph embedding model \(M\), we
used a classifier \(g_M(Y) = \sigma(Y \cdot Y^T)\), where
\(\sigma(x) = \frac{1}{1 + e^{-x}}\). The performance of the classifier
\(g_M\) was measured with the area under the receiver-operating curve
(ROC AUC). To report a signed difference in ROC AUC scores for two
classifiers \(g_M, g_{M'}\), we used the following metric
\(\phi(g_M, g_{M'}) = \frac{\rocaucscore(g_M) - \rocaucscore(g_{M'})}{\rocaucscore(g_{M'})}.\)

Overall, we made sure that there was no double-dipping into the data,
and that no information leakage from the test set took place. For each
graph embedding model, to determine the best model parameters, we let
the Bayesian hyperparameter optimization pipeline run for 50 trials on
the fraction of the original training set (80\%). Model selection was
performed by optimizing the ROC AUC score on the other fraction (20\%)
of the original training set, commonly referred as the validation set.
Best selected model parameters were then used in the final link
prediction evaluation on the holdout test set. Before the evaluation,
each graph embedding model was retrained on the full training set using
the optimized parameters from the optimization pipeline.

\hypertarget{sec:results}{%
\section{Results}\label{sec:results}}

\hypertarget{link-prediction}{%
\subsection{Link prediction}\label{link-prediction}}

\begin{figure}[!h]
    \begin{center}
        \includegraphics[width=1.\linewidth]{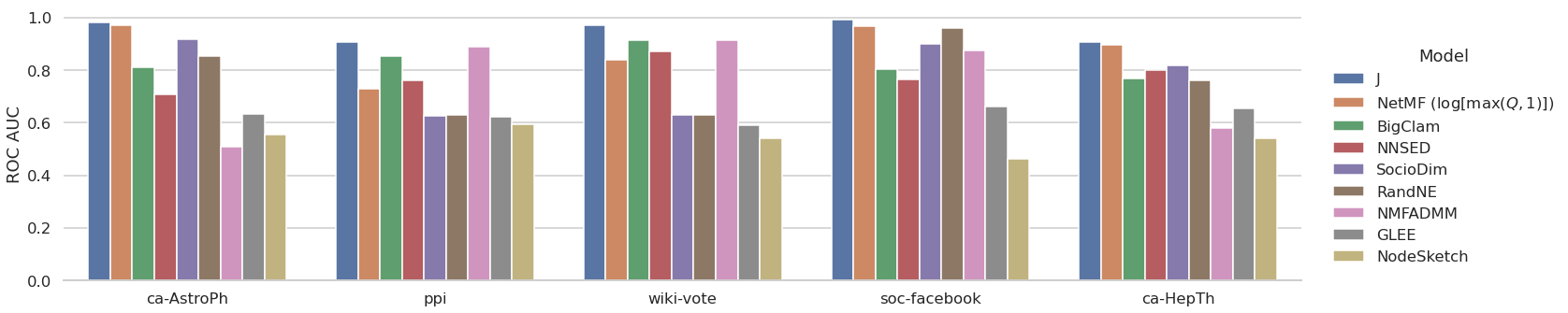}
    \end{center}
    \caption{\label{fig:test-roc-auc}ROC AUC scores on the final holdout test set.}
\end{figure}

Our approach outperformed all the considered models on all graphs, as
documented by the highest ROC AUC scores on the holdout test set (Table
\ref{tab:comparison-with-baselines}, Figure \ref{fig:test-roc-auc}).
Compared to NetMF (\(\truncatelogq\)) the proposed approach, \(J\),
increased the ROC AUC score by 24.2\% and 15.6\% on the graphs wiki-vote
and ppi. The improvement of link prediction performance on ca-AstroPh,
ca-HepTh, and soc-facebook were 1.2\%, 1.4\%, and 2.8\%, respectively.
BigClam, NNSED, and NMFADMM models had better ROC AUC scores than NetMF
(\(\truncatelogq\)) on wiki-vote and ppi. On the remaining graphs, the
baseline version of NetMF outperformed all other state-of-the-art link
prediction models (except ours).

\begin{widetable}[\label{tab:comparison-with-baselines} Link prediction performance comparison between the proposed approach and baselines. Following a Bayesian hyperparameter optimization consisting of 50 trials on the training set, the best performing model was selected and its ROC AUC score on the holdout test is reported; the best ROC AUC scores in \textbf{bold}.]

\begin{longtable}[]{@{}llllll@{}}
\toprule
\begin{minipage}[b]{0.21\columnwidth}\raggedright
Model \(M\)\strut
\end{minipage} & \begin{minipage}[b]{0.12\columnwidth}\raggedright
ca-AstroPh\strut
\end{minipage} & \begin{minipage}[b]{0.12\columnwidth}\raggedright
ca-HepTh\strut
\end{minipage} & \begin{minipage}[b]{0.12\columnwidth}\raggedright
ppi\strut
\end{minipage} & \begin{minipage}[b]{0.12\columnwidth}\raggedright
soc-facebook\strut
\end{minipage} & \begin{minipage}[b]{0.12\columnwidth}\raggedright
wiki-vote\strut
\end{minipage}\tabularnewline
\midrule
\endhead
\begin{minipage}[t]{0.21\columnwidth}\raggedright
\textbf{baseline}\strut
\end{minipage} & \begin{minipage}[t]{0.12\columnwidth}\raggedright
\strut
\end{minipage} & \begin{minipage}[t]{0.12\columnwidth}\raggedright
\strut
\end{minipage} & \begin{minipage}[t]{0.12\columnwidth}\raggedright
\strut
\end{minipage} & \begin{minipage}[t]{0.12\columnwidth}\raggedright
\strut
\end{minipage} & \begin{minipage}[t]{0.12\columnwidth}\raggedright
\strut
\end{minipage}\tabularnewline
\begin{minipage}[t]{0.21\columnwidth}\raggedright
NetMF (\(\truncatelogq\))\strut
\end{minipage} & \begin{minipage}[t]{0.12\columnwidth}\raggedright
97.0 (0.0\%)\strut
\end{minipage} & \begin{minipage}[t]{0.12\columnwidth}\raggedright
89.4 (0.0\%)\strut
\end{minipage} & \begin{minipage}[t]{0.12\columnwidth}\raggedright
73.0 (0.0\%)\strut
\end{minipage} & \begin{minipage}[t]{0.12\columnwidth}\raggedright
96.6 (0.0\%)\strut
\end{minipage} & \begin{minipage}[t]{0.12\columnwidth}\raggedright
83.9 (0.0\%)\strut
\end{minipage}\tabularnewline
\begin{minipage}[t]{0.21\columnwidth}\raggedright
------------------------\strut
\end{minipage} & \begin{minipage}[t]{0.12\columnwidth}\raggedright
--------------\strut
\end{minipage} & \begin{minipage}[t]{0.12\columnwidth}\raggedright
--------------\strut
\end{minipage} & \begin{minipage}[t]{0.12\columnwidth}\raggedright
--------------\strut
\end{minipage} & \begin{minipage}[t]{0.12\columnwidth}\raggedright
--------------\strut
\end{minipage} & \begin{minipage}[t]{0.12\columnwidth}\raggedright
--------------\strut
\end{minipage}\tabularnewline
\begin{minipage}[t]{0.21\columnwidth}\raggedright
\textbf{proposed}\strut
\end{minipage} & \begin{minipage}[t]{0.12\columnwidth}\raggedright
\strut
\end{minipage} & \begin{minipage}[t]{0.12\columnwidth}\raggedright
\strut
\end{minipage} & \begin{minipage}[t]{0.12\columnwidth}\raggedright
\strut
\end{minipage} & \begin{minipage}[t]{0.12\columnwidth}\raggedright
\strut
\end{minipage} & \begin{minipage}[t]{0.12\columnwidth}\raggedright
\strut
\end{minipage}\tabularnewline
\begin{minipage}[t]{0.21\columnwidth}\raggedright
J\strut
\end{minipage} & \begin{minipage}[t]{0.12\columnwidth}\raggedright
\textbf{98.2 (1.2\%)}\strut
\end{minipage} & \begin{minipage}[t]{0.12\columnwidth}\raggedright
\textbf{90.7 (1.4\%)}\strut
\end{minipage} & \begin{minipage}[t]{0.12\columnwidth}\raggedright
\textbf{90.7 (24.2\%)}\strut
\end{minipage} & \begin{minipage}[t]{0.12\columnwidth}\raggedright
\textbf{99.3 (2.8\%)}\strut
\end{minipage} & \begin{minipage}[t]{0.12\columnwidth}\raggedright
\textbf{96.9 (15.6\%)}\strut
\end{minipage}\tabularnewline
\begin{minipage}[t]{0.21\columnwidth}\raggedright
------------------------\strut
\end{minipage} & \begin{minipage}[t]{0.12\columnwidth}\raggedright
--------------\strut
\end{minipage} & \begin{minipage}[t]{0.12\columnwidth}\raggedright
--------------\strut
\end{minipage} & \begin{minipage}[t]{0.12\columnwidth}\raggedright
--------------\strut
\end{minipage} & \begin{minipage}[t]{0.12\columnwidth}\raggedright
--------------\strut
\end{minipage} & \begin{minipage}[t]{0.12\columnwidth}\raggedright
--------------\strut
\end{minipage}\tabularnewline
\begin{minipage}[t]{0.21\columnwidth}\raggedright
\textbf{matrix-factorization-based}\strut
\end{minipage} & \begin{minipage}[t]{0.12\columnwidth}\raggedright
\strut
\end{minipage} & \begin{minipage}[t]{0.12\columnwidth}\raggedright
\strut
\end{minipage} & \begin{minipage}[t]{0.12\columnwidth}\raggedright
\strut
\end{minipage} & \begin{minipage}[t]{0.12\columnwidth}\raggedright
\strut
\end{minipage} & \begin{minipage}[t]{0.12\columnwidth}\raggedright
\strut
\end{minipage}\tabularnewline
\begin{minipage}[t]{0.21\columnwidth}\raggedright
BigClam\strut
\end{minipage} & \begin{minipage}[t]{0.12\columnwidth}\raggedright
81.2 (-16.3\%)\strut
\end{minipage} & \begin{minipage}[t]{0.12\columnwidth}\raggedright
77.0 (-14.0\%)\strut
\end{minipage} & \begin{minipage}[t]{0.12\columnwidth}\raggedright
85.4 (16.9\%)\strut
\end{minipage} & \begin{minipage}[t]{0.12\columnwidth}\raggedright
80.4 (-16.8\%)\strut
\end{minipage} & \begin{minipage}[t]{0.12\columnwidth}\raggedright
91.5 (9.1\%)\strut
\end{minipage}\tabularnewline
\begin{minipage}[t]{0.21\columnwidth}\raggedright
GLEE\strut
\end{minipage} & \begin{minipage}[t]{0.12\columnwidth}\raggedright
63.2 (-34.8\%)\strut
\end{minipage} & \begin{minipage}[t]{0.12\columnwidth}\raggedright
65.4 (-26.8\%)\strut
\end{minipage} & \begin{minipage}[t]{0.12\columnwidth}\raggedright
62.1 (-14.9\%)\strut
\end{minipage} & \begin{minipage}[t]{0.12\columnwidth}\raggedright
66.2 (-31.5\%)\strut
\end{minipage} & \begin{minipage}[t]{0.12\columnwidth}\raggedright
59.1 (-29.5\%)\strut
\end{minipage}\tabularnewline
\begin{minipage}[t]{0.21\columnwidth}\raggedright
NMFADMM\strut
\end{minipage} & \begin{minipage}[t]{0.12\columnwidth}\raggedright
50.8 (-47.7\%)\strut
\end{minipage} & \begin{minipage}[t]{0.12\columnwidth}\raggedright
57.8 (-35.4\%)\strut
\end{minipage} & \begin{minipage}[t]{0.12\columnwidth}\raggedright
89.0 (21.9\%)\strut
\end{minipage} & \begin{minipage}[t]{0.12\columnwidth}\raggedright
87.5 (-9.4\%)\strut
\end{minipage} & \begin{minipage}[t]{0.12\columnwidth}\raggedright
91.4 (9.0\%)\strut
\end{minipage}\tabularnewline
\begin{minipage}[t]{0.21\columnwidth}\raggedright
NNSED\strut
\end{minipage} & \begin{minipage}[t]{0.12\columnwidth}\raggedright
70.8 (-27.0\%)\strut
\end{minipage} & \begin{minipage}[t]{0.12\columnwidth}\raggedright
79.9 (-10.6\%)\strut
\end{minipage} & \begin{minipage}[t]{0.12\columnwidth}\raggedright
76.0 (4.2\%)\strut
\end{minipage} & \begin{minipage}[t]{0.12\columnwidth}\raggedright
76.6 (-20.8\%)\strut
\end{minipage} & \begin{minipage}[t]{0.12\columnwidth}\raggedright
87.0 (3.8\%)\strut
\end{minipage}\tabularnewline
\begin{minipage}[t]{0.21\columnwidth}\raggedright
NodeSketch\strut
\end{minipage} & \begin{minipage}[t]{0.12\columnwidth}\raggedright
55.6 (-42.8\%)\strut
\end{minipage} & \begin{minipage}[t]{0.12\columnwidth}\raggedright
54.1 (-39.5\%)\strut
\end{minipage} & \begin{minipage}[t]{0.12\columnwidth}\raggedright
59.3 (-18.8\%)\strut
\end{minipage} & \begin{minipage}[t]{0.12\columnwidth}\raggedright
46.2 (-52.2\%)\strut
\end{minipage} & \begin{minipage}[t]{0.12\columnwidth}\raggedright
54.2 (-35.4\%)\strut
\end{minipage}\tabularnewline
\begin{minipage}[t]{0.21\columnwidth}\raggedright
RandNE\strut
\end{minipage} & \begin{minipage}[t]{0.12\columnwidth}\raggedright
85.2 (-12.2\%)\strut
\end{minipage} & \begin{minipage}[t]{0.12\columnwidth}\raggedright
76.1 (-14.9\%)\strut
\end{minipage} & \begin{minipage}[t]{0.12\columnwidth}\raggedright
62.8 (-13.9\%)\strut
\end{minipage} & \begin{minipage}[t]{0.12\columnwidth}\raggedright
96.0 (-0.7\%)\strut
\end{minipage} & \begin{minipage}[t]{0.12\columnwidth}\raggedright
63.1 (-24.7\%)\strut
\end{minipage}\tabularnewline
\begin{minipage}[t]{0.21\columnwidth}\raggedright
SocioDim\strut
\end{minipage} & \begin{minipage}[t]{0.12\columnwidth}\raggedright
91.6 (-5.6\%)\strut
\end{minipage} & \begin{minipage}[t]{0.12\columnwidth}\raggedright
81.9 (-8.5\%)\strut
\end{minipage} & \begin{minipage}[t]{0.12\columnwidth}\raggedright
62.7 (-14.2\%)\strut
\end{minipage} & \begin{minipage}[t]{0.12\columnwidth}\raggedright
90.1 (-6.8\%)\strut
\end{minipage} & \begin{minipage}[t]{0.12\columnwidth}\raggedright
62.9 (-25.0\%)\strut
\end{minipage}\tabularnewline
\bottomrule
\end{longtable}

\end{widetable}

\hypertarget{truncating-or-smoothing-low-co-occurrence-random-walk-pairs}{%
\subsubsection{Truncating or smoothing low co-occurrence random walk
pairs?}\label{truncating-or-smoothing-low-co-occurrence-random-walk-pairs}}

The original version of NetMF, \(\truncatelogq\) truncates the
contribution of low-occurence pairs due to the \(\log [\max(\cdot, 1)]\)
transformation of the matrix \(Q\). Instead, we considered a version of
NetMF, \(\sigmoidq\), that smoothens these entries, instead of
truncating them, and compared its best-optimized model to the best
optimized \(\truncatelogq\) model. To smoothen entries in the \(Q\)
matrix, we used the sigmoid function,
\(\sigma(x) = \frac{1}{1 + e^{-x}}\), which squishes inputs from the
\([-\infty, \infty]\) into the \([0, 1]\) range. Due to this
transformation the information from unobserved pairs, or the pairs with
very low joint probability, is not lost. However, this comes at the cost
of decomposing a dense matrix to get the graph embeddings \(Y\). Note
that we cannot decompose the original shifted PMI matrix \(\logq\),
since it leads to decomposing ill-defined \(\log 0 = - \infty\) entries.
The NetMF version that smoothens low-occurrence pairs outperformed the
baseline on wiki-vote and ppi, had the same performance on soc-facebook,
and obtained a slightly worse performance on ca-AstroPh and ca-HepTh
graphs. Table \ref{tab:truncate-or-smoothen} shows head two head
comparison of two versions' ROC AUC scores.

\begin{widetable}[\label{tab:truncate-or-smoothen} Comparison of test ROC AUC scores for two versions of NetMF algorithm: original version NetMF ($\truncatelogq$) that truncates node pairs with low PMI, and NetMF ($\sigmoidq$) that smoothens low PMI pairs. Best scores are in \textbf{bold}.]

\begin{longtable}[]{@{}llllll@{}}
\toprule
\begin{minipage}[b]{0.21\columnwidth}\raggedright
NetMF version\strut
\end{minipage} & \begin{minipage}[b]{0.12\columnwidth}\raggedright
ca-AstroPh\strut
\end{minipage} & \begin{minipage}[b]{0.12\columnwidth}\raggedright
ca-HepTh\strut
\end{minipage} & \begin{minipage}[b]{0.12\columnwidth}\raggedright
ppi\strut
\end{minipage} & \begin{minipage}[b]{0.12\columnwidth}\raggedright
soc-facebook\strut
\end{minipage} & \begin{minipage}[b]{0.12\columnwidth}\raggedright
wiki-vote\strut
\end{minipage}\tabularnewline
\midrule
\endhead
\begin{minipage}[t]{0.21\columnwidth}\raggedright
\textbf{truncate}\strut
\end{minipage} & \begin{minipage}[t]{0.12\columnwidth}\raggedright
\strut
\end{minipage} & \begin{minipage}[t]{0.12\columnwidth}\raggedright
\strut
\end{minipage} & \begin{minipage}[t]{0.12\columnwidth}\raggedright
\strut
\end{minipage} & \begin{minipage}[t]{0.12\columnwidth}\raggedright
\strut
\end{minipage} & \begin{minipage}[t]{0.12\columnwidth}\raggedright
\strut
\end{minipage}\tabularnewline
\begin{minipage}[t]{0.21\columnwidth}\raggedright
NetMF (\(\truncatelogq\))\strut
\end{minipage} & \begin{minipage}[t]{0.12\columnwidth}\raggedright
\textbf{97.0 (0.0\%)}\strut
\end{minipage} & \begin{minipage}[t]{0.12\columnwidth}\raggedright
\textbf{89.4 (0.0\%)}\strut
\end{minipage} & \begin{minipage}[t]{0.12\columnwidth}\raggedright
73.0 (0.0\%)\strut
\end{minipage} & \begin{minipage}[t]{0.12\columnwidth}\raggedright
\textbf{96.6 (0.0\%)}\strut
\end{minipage} & \begin{minipage}[t]{0.12\columnwidth}\raggedright
83.9 (0.0\%)\strut
\end{minipage}\tabularnewline
\begin{minipage}[t]{0.21\columnwidth}\raggedright
\textbf{smoothen}\strut
\end{minipage} & \begin{minipage}[t]{0.12\columnwidth}\raggedright
\strut
\end{minipage} & \begin{minipage}[t]{0.12\columnwidth}\raggedright
\strut
\end{minipage} & \begin{minipage}[t]{0.12\columnwidth}\raggedright
\strut
\end{minipage} & \begin{minipage}[t]{0.12\columnwidth}\raggedright
\strut
\end{minipage} & \begin{minipage}[t]{0.12\columnwidth}\raggedright
\strut
\end{minipage}\tabularnewline
\begin{minipage}[t]{0.21\columnwidth}\raggedright
NetMF (\(\sigmoidq\))\strut
\end{minipage} & \begin{minipage}[t]{0.12\columnwidth}\raggedright
96.9 (-0.2\%)\strut
\end{minipage} & \begin{minipage}[t]{0.12\columnwidth}\raggedright
88.7 (-0.9\%)\strut
\end{minipage} & \begin{minipage}[t]{0.12\columnwidth}\raggedright
\textbf{84.3 (15.5\%)}\strut
\end{minipage} & \begin{minipage}[t]{0.12\columnwidth}\raggedright
\textbf{96.6 (0.0\%)}\strut
\end{minipage} & \begin{minipage}[t]{0.12\columnwidth}\raggedright
\textbf{90.4 (7.8\%)}\strut
\end{minipage}\tabularnewline
\bottomrule
\end{longtable}

\end{widetable}

\hypertarget{generalization-gap}{%
\subsection{Generalization gap}\label{generalization-gap}}

\begin{figure}[!h]
    \begin{center}
        \includegraphics[width=1.\linewidth]{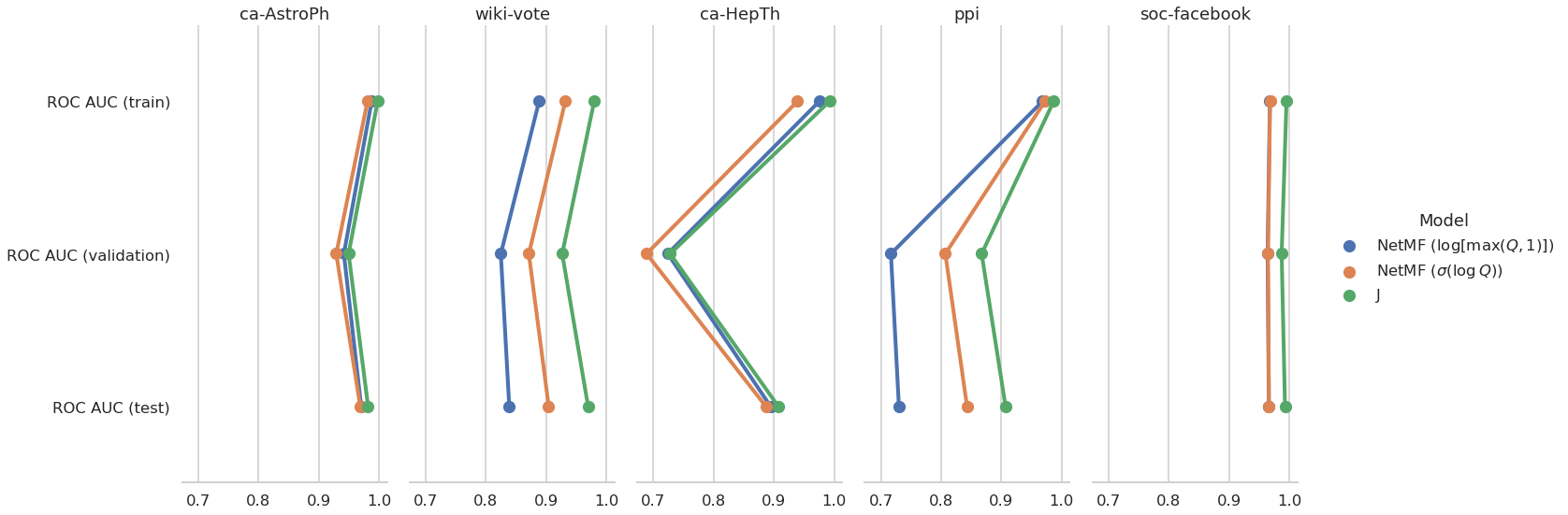}
    \end{center}
    \caption{\label{fig:generalization-gap-roc-auc}Generalization gap between the train, validation, and test ROC AUC scores.}
\end{figure}

We looked into the generalization capability of \(J\) and baseline NetMF
models, NetMF(\(\truncatelogq\)) and NetMF(\(\sigmoidq\)). In the
statistical learning theory, under the bias-variance tradeoff paradigm,
we expect the test score to be slightly worse than the training score.
Moreover, really big negative differences may point to the overfitting
phenomenon, where a prediction model has learned the training dataset
but failed to generalize on the test set. Figure
\ref{fig:generalization-gap-roc-auc} plots train, validation, and test
ROC AUC scores for the proposed approach and two versions of NetMF.
Compared to the baseline, the proposed approach had better absolute
scores on all graphs, and specifically on the graphs wiki-vote and ppi.
On these two graphs, the test ROC AUC score with respect to the train
ROC AUC score for the approach of Qiu et al.~\citep{qiu2018} dropped 6\%
and 32\%, respectively. For comparison, it only dropped 1\% and 9\% for
the proposed approach. For these three approaches, it is interesting to
note that the graph soc-facebook had the best generalization gap, as
differences between the ROC AUC scores were less than 1\%. The graph
ca-HepTh had the biggest difference between the validation and the train
ROC AUC score, dropping \textasciitilde36\% for all three models,
however, the gaps between the train and test scores were much smaller
(5-9\%).

\hypertarget{optimization-duration-and-parameter-sensitivity-analysis}{%
\subsection{Optimization duration and parameter sensitivity
analysis}\label{optimization-duration-and-parameter-sensitivity-analysis}}

Table \ref{tab:optimization-duration} documents the optimization
duration for all the considered models. NodeSketch was the most
computationally intensive model to optimize, RandNE was the fastest
model to optimize. Two NetMF versions and the proposed approach took
more time to optimize than the median optimization duration. Notably,
the sigmoid version of NetMF (\(\sigmoidq\)) took less time to optimize
than the original version that used the truncated log transformation
(\(\truncatelogq\)). The optimization duration for the proposed approach
\(J\) and the original NetMF was comparable.

\begin{widetable}[\label{tab:optimization-duration}Total optimization duration for 50 trials in minutes, average trial optimization duration in seconds (in parenthesis).]

\begin{longtable}[]{@{}llllll@{}}
\toprule
\begin{minipage}[b]{0.20\columnwidth}\raggedright
Graph embedding model\strut
\end{minipage} & \begin{minipage}[b]{0.13\columnwidth}\raggedright
ca-AstroPh\strut
\end{minipage} & \begin{minipage}[b]{0.11\columnwidth}\raggedright
ca-HepTh\strut
\end{minipage} & \begin{minipage}[b]{0.14\columnwidth}\raggedright
ppi\strut
\end{minipage} & \begin{minipage}[b]{0.13\columnwidth}\raggedright
soc-facebook\strut
\end{minipage} & \begin{minipage}[b]{0.13\columnwidth}\raggedright
wiki-vote\strut
\end{minipage}\tabularnewline
\midrule
\endhead
\begin{minipage}[t]{0.20\columnwidth}\raggedright
\textbf{baseline}\strut
\end{minipage} & \begin{minipage}[t]{0.13\columnwidth}\raggedright
\strut
\end{minipage} & \begin{minipage}[t]{0.11\columnwidth}\raggedright
\strut
\end{minipage} & \begin{minipage}[t]{0.14\columnwidth}\raggedright
\strut
\end{minipage} & \begin{minipage}[t]{0.13\columnwidth}\raggedright
\strut
\end{minipage} & \begin{minipage}[t]{0.13\columnwidth}\raggedright
\strut
\end{minipage}\tabularnewline
\begin{minipage}[t]{0.20\columnwidth}\raggedright
NetMF (\(\truncatelogq\))\strut
\end{minipage} & \begin{minipage}[t]{0.13\columnwidth}\raggedright
201 m (241 s)\strut
\end{minipage} & \begin{minipage}[t]{0.11\columnwidth}\raggedright
32 m (38 s)\strut
\end{minipage} & \begin{minipage}[t]{0.14\columnwidth}\raggedright
11 m (13 s)\strut
\end{minipage} & \begin{minipage}[t]{0.13\columnwidth}\raggedright
5 m (7 s)\strut
\end{minipage} & \begin{minipage}[t]{0.13\columnwidth}\raggedright
8 m (10 s)\strut
\end{minipage}\tabularnewline
\begin{minipage}[t]{0.20\columnwidth}\raggedright
\textbf{proposed}\strut
\end{minipage} & \begin{minipage}[t]{0.13\columnwidth}\raggedright
\strut
\end{minipage} & \begin{minipage}[t]{0.11\columnwidth}\raggedright
\strut
\end{minipage} & \begin{minipage}[t]{0.14\columnwidth}\raggedright
\strut
\end{minipage} & \begin{minipage}[t]{0.13\columnwidth}\raggedright
\strut
\end{minipage} & \begin{minipage}[t]{0.13\columnwidth}\raggedright
\strut
\end{minipage}\tabularnewline
\begin{minipage}[t]{0.20\columnwidth}\raggedright
J\strut
\end{minipage} & \begin{minipage}[t]{0.13\columnwidth}\raggedright
180 m (216 s)\strut
\end{minipage} & \begin{minipage}[t]{0.11\columnwidth}\raggedright
42 m (51 s)\strut
\end{minipage} & \begin{minipage}[t]{0.14\columnwidth}\raggedright
4 m (5 s)\strut
\end{minipage} & \begin{minipage}[t]{0.13\columnwidth}\raggedright
13 m (15 s)\strut
\end{minipage} & \begin{minipage}[t]{0.13\columnwidth}\raggedright
23 m (28 s)\strut
\end{minipage}\tabularnewline
\begin{minipage}[t]{0.20\columnwidth}\raggedright
NetMF (\(\sigmoidq\))\strut
\end{minipage} & \begin{minipage}[t]{0.13\columnwidth}\raggedright
96 m (115 s)\strut
\end{minipage} & \begin{minipage}[t]{0.11\columnwidth}\raggedright
14 m (16 s)\strut
\end{minipage} & \begin{minipage}[t]{0.14\columnwidth}\raggedright
2 m (2 s)\strut
\end{minipage} & \begin{minipage}[t]{0.13\columnwidth}\raggedright
2 m (3 s)\strut
\end{minipage} & \begin{minipage}[t]{0.13\columnwidth}\raggedright
9 m (10 s)\strut
\end{minipage}\tabularnewline
\begin{minipage}[t]{0.20\columnwidth}\raggedright
\textbf{matrix-factorization-based}\strut
\end{minipage} & \begin{minipage}[t]{0.13\columnwidth}\raggedright
\strut
\end{minipage} & \begin{minipage}[t]{0.11\columnwidth}\raggedright
\strut
\end{minipage} & \begin{minipage}[t]{0.14\columnwidth}\raggedright
\strut
\end{minipage} & \begin{minipage}[t]{0.13\columnwidth}\raggedright
\strut
\end{minipage} & \begin{minipage}[t]{0.13\columnwidth}\raggedright
\strut
\end{minipage}\tabularnewline
\begin{minipage}[t]{0.20\columnwidth}\raggedright
BigClam\strut
\end{minipage} & \begin{minipage}[t]{0.13\columnwidth}\raggedright
27 m (33 s)\strut
\end{minipage} & \begin{minipage}[t]{0.11\columnwidth}\raggedright
8 m (10 s)\strut
\end{minipage} & \begin{minipage}[t]{0.14\columnwidth}\raggedright
4 m (5 s)\strut
\end{minipage} & \begin{minipage}[t]{0.13\columnwidth}\raggedright
6 m (7 s)\strut
\end{minipage} & \begin{minipage}[t]{0.13\columnwidth}\raggedright
15 m (18 s)\strut
\end{minipage}\tabularnewline
\begin{minipage}[t]{0.20\columnwidth}\raggedright
GLEE\strut
\end{minipage} & \begin{minipage}[t]{0.13\columnwidth}\raggedright
78 m (94 s)\strut
\end{minipage} & \begin{minipage}[t]{0.11\columnwidth}\raggedright
15 m (18 s)\strut
\end{minipage} & \begin{minipage}[t]{0.14\columnwidth}\raggedright
4 m (5 s)\strut
\end{minipage} & \begin{minipage}[t]{0.13\columnwidth}\raggedright
6 m (7 s)\strut
\end{minipage} & \begin{minipage}[t]{0.13\columnwidth}\raggedright
9 m (11 s)\strut
\end{minipage}\tabularnewline
\begin{minipage}[t]{0.20\columnwidth}\raggedright
NMFADMM\strut
\end{minipage} & \begin{minipage}[t]{0.13\columnwidth}\raggedright
56 m (67 s)\strut
\end{minipage} & \begin{minipage}[t]{0.11\columnwidth}\raggedright
73 m (88 s)\strut
\end{minipage} & \begin{minipage}[t]{0.14\columnwidth}\raggedright
26 m (31 s)\strut
\end{minipage} & \begin{minipage}[t]{0.13\columnwidth}\raggedright
61 m (73 s)\strut
\end{minipage} & \begin{minipage}[t]{0.13\columnwidth}\raggedright
40 m (48 s)\strut
\end{minipage}\tabularnewline
\begin{minipage}[t]{0.20\columnwidth}\raggedright
NNSED\strut
\end{minipage} & \begin{minipage}[t]{0.13\columnwidth}\raggedright
21 m (25 s)\strut
\end{minipage} & \begin{minipage}[t]{0.11\columnwidth}\raggedright
5 m (6 s)\strut
\end{minipage} & \begin{minipage}[t]{0.14\columnwidth}\raggedright
2 m (2 s)\strut
\end{minipage} & \begin{minipage}[t]{0.13\columnwidth}\raggedright
2 m (2 s)\strut
\end{minipage} & \begin{minipage}[t]{0.13\columnwidth}\raggedright
5 m (6 s)\strut
\end{minipage}\tabularnewline
\begin{minipage}[t]{0.20\columnwidth}\raggedright
NodeSketch\strut
\end{minipage} & \begin{minipage}[t]{0.13\columnwidth}\raggedright
3475 m (4170 s)\strut
\end{minipage} & \begin{minipage}[t]{0.11\columnwidth}\raggedright
174 m (209 s)\strut
\end{minipage} & \begin{minipage}[t]{0.14\columnwidth}\raggedright
1899 m (2279 s)\strut
\end{minipage} & \begin{minipage}[t]{0.13\columnwidth}\raggedright
3594 m (4313 s)\strut
\end{minipage} & \begin{minipage}[t]{0.13\columnwidth}\raggedright
4445 m (5334 s)\strut
\end{minipage}\tabularnewline
\begin{minipage}[t]{0.20\columnwidth}\raggedright
RandNE\strut
\end{minipage} & \begin{minipage}[t]{0.13\columnwidth}\raggedright
6 m (8 s)\strut
\end{minipage} & \begin{minipage}[t]{0.11\columnwidth}\raggedright
1 m (1 s)\strut
\end{minipage} & \begin{minipage}[t]{0.14\columnwidth}\raggedright
\textless1 m (\textless1 s)\strut
\end{minipage} & \begin{minipage}[t]{0.13\columnwidth}\raggedright
\textless1 m (\textless1 s)\strut
\end{minipage} & \begin{minipage}[t]{0.13\columnwidth}\raggedright
1 m (1 s)\strut
\end{minipage}\tabularnewline
\begin{minipage}[t]{0.20\columnwidth}\raggedright
SocioDim\strut
\end{minipage} & \begin{minipage}[t]{0.13\columnwidth}\raggedright
50 m (60 s)\strut
\end{minipage} & \begin{minipage}[t]{0.11\columnwidth}\raggedright
17 m (20 s)\strut
\end{minipage} & \begin{minipage}[t]{0.14\columnwidth}\raggedright
3 m (4 s)\strut
\end{minipage} & \begin{minipage}[t]{0.13\columnwidth}\raggedright
4 m (4 s)\strut
\end{minipage} & \begin{minipage}[t]{0.13\columnwidth}\raggedright
8 m (10 s)\strut
\end{minipage}\tabularnewline
\begin{minipage}[t]{0.20\columnwidth}\raggedright
-------------------------\strut
\end{minipage} & \begin{minipage}[t]{0.13\columnwidth}\raggedright
\strut
\end{minipage} & \begin{minipage}[t]{0.11\columnwidth}\raggedright
\strut
\end{minipage} & \begin{minipage}[t]{0.14\columnwidth}\raggedright
\strut
\end{minipage} & \begin{minipage}[t]{0.13\columnwidth}\raggedright
\strut
\end{minipage} & \begin{minipage}[t]{0.13\columnwidth}\raggedright
\strut
\end{minipage}\tabularnewline
\begin{minipage}[t]{0.20\columnwidth}\raggedright
\textbf{median for each graph}\strut
\end{minipage} & \begin{minipage}[t]{0.13\columnwidth}\raggedright
\strut
\end{minipage} & \begin{minipage}[t]{0.11\columnwidth}\raggedright
\strut
\end{minipage} & \begin{minipage}[t]{0.14\columnwidth}\raggedright
\strut
\end{minipage} & \begin{minipage}[t]{0.13\columnwidth}\raggedright
\strut
\end{minipage} & \begin{minipage}[t]{0.13\columnwidth}\raggedright
\strut
\end{minipage}\tabularnewline
\begin{minipage}[t]{0.20\columnwidth}\raggedright
\strut
\end{minipage} & \begin{minipage}[t]{0.13\columnwidth}\raggedright
96 m (115 s)\strut
\end{minipage} & \begin{minipage}[t]{0.11\columnwidth}\raggedright
17 m (20 s)\strut
\end{minipage} & \begin{minipage}[t]{0.14\columnwidth}\raggedright
5 m (6 s)\strut
\end{minipage} & \begin{minipage}[t]{0.13\columnwidth}\raggedright
6 m (7 s)\strut
\end{minipage} & \begin{minipage}[t]{0.13\columnwidth}\raggedright
15 m (18 s)\strut
\end{minipage}\tabularnewline
\bottomrule
\end{longtable}

\end{widetable}

\begin{figure}[!h]
    \begin{center}
        \includegraphics[width=1.\linewidth]{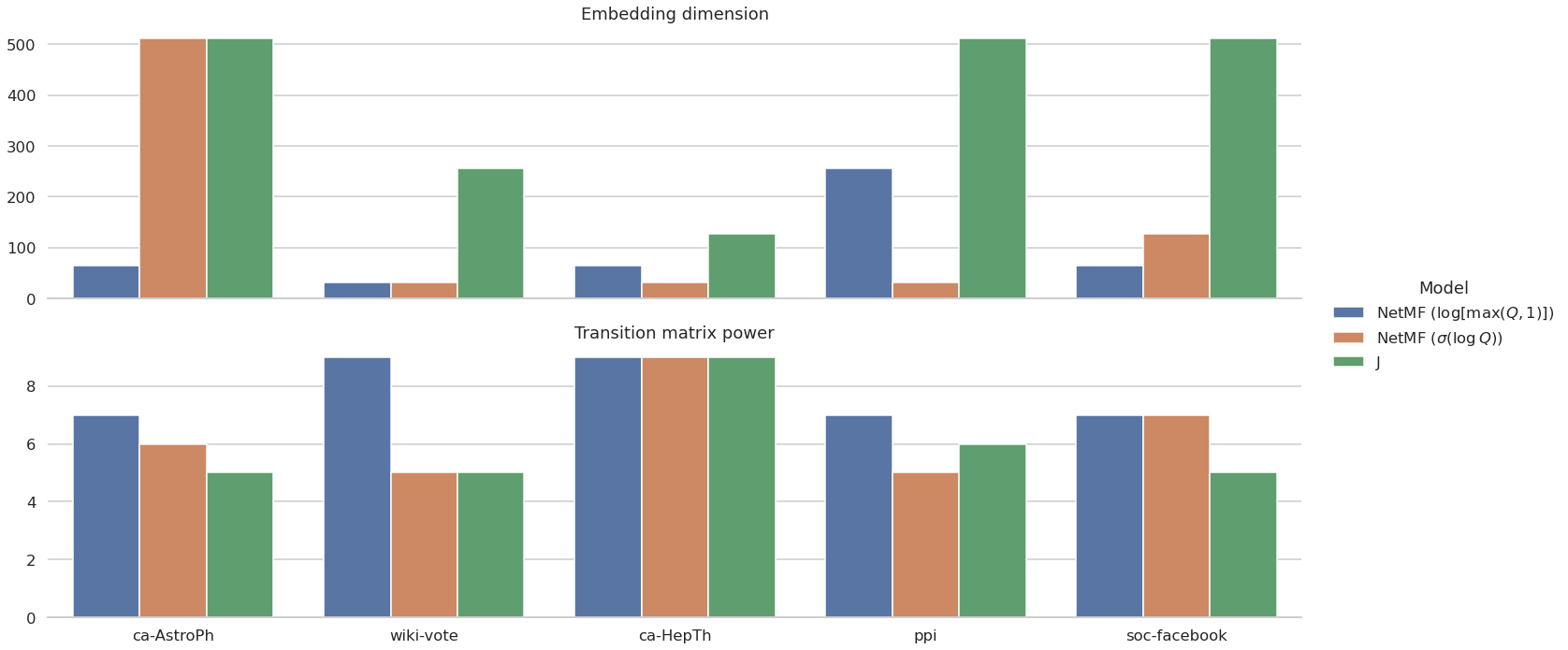}
    \end{center}
    \caption{\label{fig:hyperparameters}Optimized hyperparameters that lead to the best ROC AUC scores.}
\end{figure}

Figure \ref{fig:hyperparameters} shows the best hyperparameters for the
proposed approach and two NetMF versions. While \(J\) required more
dimensions, i.e., bigger representation capacity, it required a smaller
random walk length. It is interesting to see that the best ROC AUC
scores were obtained when only 1 negative sample was taken into account
for the two versions of the NetMF algorithm.

\hypertarget{comparison-with-graph-embedding-models-based-on-neural-networks}{%
\subsection{Comparison with graph embedding models based on neural
networks}\label{comparison-with-graph-embedding-models-based-on-neural-networks}}

We compared our approach with graph embedding models based on neural
networks, using published results in Abu-El-Haija et al.~\citep{abu2018}
(Table~\ref{tab:comparison-with-watch}). In their work,~the authors used
a neural graph embedding model WatchYourStep that was implicitly
factorizing the expected values of the matrix of co-occurrence counts
\(\mathbb{E}[C]\), where each entry \(\mathbb{E}[C]_{wc}\) represents
the expected amount of times that nodes \(w, c\) will co-occur in all
random walks \citep{abu2018}. In addition, it was reported that
WatchYourStep largely outperformed other graph embedding models,
including DNGR \citep{cao2016}, node2vec \citep{grover2016}, and Assym
Proj \citep{abu2017}. We have used exactly the same training and test
splits as in \citep{abu2018} and found that our approach also
outperformed DNGR, node2vec and Assym Proj. Compared to WatchYourStep
head-to-head, our proposed method, \(J\), obtained comparable scores
(within \textasciitilde3\% ROC AUC scores), and on the graph wiki-vote
even outperformed it by \textgreater3\%.

\begin{widetable}[\label{tab:comparison-with-watch} Comparison of test ROC AUC scores of the proposed approach and the scores obtained with state-of-the-art neural graph embedding algorithms as reported in \cite{abu2018}; $^\dagger$ best scores per model. Train and test splits are exactly those from \cite{abu2018}. Overall best ROC AUC scores in \textbf{bold}, second best in \textit{italic}.]

\begin{longtable}[]{@{}llllll@{}}
\toprule
Graph embedding model & ca-AstroPh & ca-HepTh & ppi & soc-facebook &
wiki-vote\tabularnewline
\midrule
\endhead
\textbf{proposed} & & & & &\tabularnewline
J & \emph{98.2} & \emph{90.7} & \emph{90.7} & \emph{99.3} &
\textbf{96.9}\tabularnewline
\textbf{neural-network-based} & & & & &\tabularnewline
WatchYourStep (\(\mathbb{E}[C])^\dagger\) \cite{abu2018} & \textbf{98.6}
& \textbf{93.9} & \textbf{91.0} & \textbf{99.5} &
\emph{93.8}\tabularnewline
DNGR \(^\dagger\) \cite{cao2016} & 96.8 & 89.7 & 76.9 & 98.4 &
59.8\tabularnewline
node2vec \(^\dagger\) \cite{grover2016} & 97.7 & 92.0 & 81.8 &
\emph{99.3} & 64.6\tabularnewline
Assym Proj \(^\dagger\) \cite{abu2017} & 95.7 & 90.3 & 83.9 & 97.4 &
91.7\tabularnewline
\bottomrule
\end{longtable}

\end{widetable}

\hypertarget{sec:discussion}{%
\section{Discussion}\label{sec:discussion}}

In the last decade, the research on learning neural graph embeddings has
been deeply influenced by the success of the word2vec algorithm. This
word embedding algorithm used neural networks to perform low-rank
factorization of the shifted pointwise mutual information matrix, where
each entry represents a linguistic collocation of two words. In this
work, we put under question whether the collocation metric is the right
fit in the realm of real-world networks. Instead, we proposed a graph
embedding approach that relies on factorizing the joint probability
matrix \(J\), where each entry encodes the probability of two nodes to
co-occur in the same context in one random walk. Benchmarked on link
prediction tasks, our approach improved the performance of baseline
graph embeddings obtained by factorizing the shifted pointwise mutual
information matrix. Depending on the network topology, the improvement
ranged from 1.2\% to 24.2\%. In the following, we discuss a few insights
from our work that could be important in designing graph embedding
algorithms.

\hypertarget{explicit-modeling-of-negative-samples-has-little-effect-on-the-quality-of-graph-embeddings}{%
\subsection{Explicit modeling of negative samples has little effect on
the quality of graph
embeddings}\label{explicit-modeling-of-negative-samples-has-little-effect-on-the-quality-of-graph-embeddings}}

Our work ponders the importance of explicit modeling of negative samples
for link prediction problems on graphs. Qiu et al.~\citep{qiu2018} have
shown that deepwalk is optimized when the inner product of a
target-context node pair \((w_t, w_c)\) is approaching
\(\text{PMI}(w_t, w_c) - \log b\). In fact, the shifted PMI approach,
which stems from natural language processing, requires the explicit term
\(\log b\), where \(b\) is the amount of sampled negative examples
during stochastic training. However, in our experiments, we saw that the
role of the negative term in the shifted PMI formula plays a limited
role. The best-optimized models for all graphs consistently chose
\(b=1\), such that the term \(\log [b=1] = 0\) did not contribute any
new information.

\hypertarget{smoothing-low-frequency-pairs-yields-better-link-prediction-performance-than-truncating}{%
\subsection{Smoothing low-frequency pairs yields better link prediction
performance than
truncating}\label{smoothing-low-frequency-pairs-yields-better-link-prediction-performance-than-truncating}}

The original NetMF algorithm truncates the contribution of node pairs
with a low shifted PMI score using a \(\log [\max (1, \cdot) ]\)
transformation. This technical choice has been done to avoid the
ill-imposed decomposition of \(\lim_{x \to 0} \log x = -\infty\) entries
in the matrix for pairs of nodes that never appear together in a random
walk. However, due to the \(\log [\max (1, \cdot) ]\) transformation the
contribution of low-frequency pairs to learning graph embeddings is
undermined. In fact, applying the sigmoid function \(\sigma(\cdot)\) to
the argument of \(\log\), i.e., \(\sigma (\log [ \cdot ] )\), yields a
comparable (\textless1\% difference in ROC AUC score), or in some cases
much better link prediction performance (7.8\% - 15.5\% improvement). We
believe that this happens because all the variance of the low-frequency
pairs -- most entries in the graph co-occurrence statistics matrix -- is
taken into account during the matrix factorization with SVD. Therefore,
we believe that \emph{smoothening} low-frequency pairs should be favored
over \emph{truncating} them for link prediction.

\hypertarget{pointwise-mutual-information-metric-and-scale-free-property-of-real-world-networks}{%
\subsection{Pointwise mutual information metric and scale-free property
of real-world
networks}\label{pointwise-mutual-information-metric-and-scale-free-property-of-real-world-networks}}

Interestingly, the joint probability matrix \(J\) and shifted PMI matrix
are extremely related, both encode co-occurrence information from the
same graph and for the same simulated random walks, so why do we see the
difference in performance? We believe that to explain this, we need to
analyze under which circumstances a relatively high joint probability of
\(i\) and \(j\) appearing together encodes more useful information for
the link prediction problem. Point-wise mutual information of a pair of
outcomes \(x, y\) from discrete random variables \(X\) and \(Y\)
quantifies the discrepancy between the probability of their coincidence
(co-occurrence) given their joint and individual distributions. In
computational linguistics, PMI is used for finding collocations and
associations between words. Good collocation pairs have high \(\pmi\)
because the probability of co-occurrence \(p(i, j)\) is only slightly
lower than the probabilities of occurrence of each word (\(p(i)\) and
\(p(j)\)). On the other hand, \(\pmi(i, j)\) is zero when these two
outcomes are independent, i.e., \(p(i, j) = p(i)p(j)\));
\(\pmi(i, j) = \log \left[ \frac{p(i,j)}{p(i)p(j)} \right] = \log 1 = 0\).
And it is undefined when \(i, j\) do not co-occur at all, i.e.,
\(\log \pmi(i, j) = \log \left[ \frac{p(i,j)}{p(i)p(j)} \right] = \log 0 = - \infty\),
when \(p(i, j) = 0\). On the other hand, \(\pmi\) is maximized when two
outcomes are perfectly associated, i.e., \(p(i | j) = 1\) or
\(p(j | i) = 1\). For instance, a pair of words \textsc{puerto rico} has
a very high \(\pmi\), since they are rarely used individually in the
\footnote{\url{https://en.wikipedia.org/wiki/Pointwise_mutual_information}}Wikipedia
articles. From the same source, a pair of words \textsc{it the}, despite
having the same joint probability value as \textsc{puerto rico}, will
have a very low \(\pmi\) score because their joint probability is small
in comparison with their individual probabilities.

Translating to the graph domain, consider two connected nodes \(i, j\),
such that both \(i\) and \(j\) are extremely frequent (very high
centrality). If we compute the \(\pmi\) of this pair it will give us a
very low score because the numerator (joint probability) is dominated by
the denominator (big individual probabilities), therefore the dot
product of embeddings for these two nodes will give us a low score,
which might be interpreted that the two nodes are not connected (while
they are). This can seriously undermine the link predictor. It is very
well known that many real-world networks, such as protein interaction,
social, and citation, networks exhibit scale-free nature, where the
degree distribution follows the power law, which results in networks
that have few hubs to which all other nodes tend to attach
\citep{barabasi2009}, \citep{dorogovtsev2002}, \citep{barabasi1999}.
Therefore, we believe that a high joint probability for two highly
frequent nodes should not be penalized, as it might be very indicative
of a very plausible link between the two nodes. Hence, we would
recommend \(J\), the joint probability matrix, as the main choice for
the computation of node embeddings for link prediction.

\hypertarget{the-importance-of-non-linear-feature-extraction-in-link-prediction}{%
\subsection{The importance of non-linear feature extraction in link
prediction}\label{the-importance-of-non-linear-feature-extraction-in-link-prediction}}

Improving the capacity of neural graph embedding models to extract
non-linear features from graph data is an active area of pattern
recognition research \citep{hu2021}. The insights from our work may help
train non-linear feature extraction link predictors more efficiently, in
particular, when it comes to hyperparameter optimization. Indeed,
non-linear neural embeddings approaches require extensive computational
costs to find an optimal choice of hyperparameters, e.g., embedding
dimension. Neural graph embedding algorithms based on explicit matrix
factorization with SVD, such as the one presented in this work, can be
used as a baseline model to determine the best embedding dimension. In
particular, the Eckart-Young-Mirsky theorem guarantees the decomposed
vectors of rank \(d\) (i.e., embedding dimension) are the best linear
approximation.

However, we should also be aware that some real-world networks may
benefit only marginally from complex non-linear feature extraction. Our
proposed linear approach obtained a similar performance as the
non-linear WatchYourStep approach by Abu-El-Haija et
al.~\citep{abu2018}.

\hypertarget{related-work}{%
\subsection{Related work}\label{related-work}}

Our work is based on the tremendous previous efforts to connect neural
embeddings for symbolic data and the solid theory of matrix
factorization~\citep{levy2014, qiu2018}. Perhaps, the most similar work
to ours, to the best of our knowledge, comes from the NLP
domain~\citep{kenyondean2019}, where they showed that the SVD
decomposition of the truncated \(\pmi\) matrix, as in~\citep{levy2014},
may worsen the quality of word embeddings.

Among the most popular approaches to learning node embeddings and
applying them to link prediction, we can outline three main classes:

\begin{itemize}
\tightlist
\item
  \textbf{Methods that are not based on neural networks}. For instance,
  BigClam \citep{yang2013bigclam}, NNSED \citep{sun2017nnsed} use
  community detection algorithms to learn node embeddings. SocioDim
  \citep{tang2009sociodim}, RandNE \citep{zhang2018randne}, NMFADMM
  \citep{sun2014nmfadmm}, NodeSketch \citep{yang2019nodesketch}, and
  GLEE \citep{torres2020} exploit neighborhood information. In our
  comparison, we found out our approach outperforms these baselines.
  Other methods that rely on feature engineering for individual nodes
  \citep{kerrache2020}, \citep{lu2015}, \citep{xu2017}, \citep{wang2017}
  have been also popular for link prediction problems. We have not
  included these methods in our comparison.
\item
  \textbf{Methods that are based on neural networks}. Deepwalk
  \citep{perozzi2014} inspired approaches include node2vec
  \citep{grover2016}, NetMF \citep{qiu2018}, DGNR \citep{cao2016}, Assym
  Proj \citep{abu2017}, and WatchYourStep \citep{abu2018}. We showed
  that the performance of our approach compares favorably.
\item
  \textbf{Methods that require per-node feature information}. Some
  algorithms require not only graph structure information, but also
  per-node feature information, e.g., gene expression values in protein
  interaction networks. Prominent examples include graph convolution
  methods for semi-supervized node classification \citep{bruna2014},
  \citep{atwood2016}, \citep{kipf2016} and graph representational
  learning using variational autoencoders \citep{wang2022}. We have not
  included these algorithms in the comparison since they require
  additional per-node feature information. It is worth mentioning that
  the predictive performance of graph convolutional neural networks can
  be significantly improved by incorporating the structural information
  about the input graph into the training process \citep{zhang2019},
  \citep{fan2020}, \citep{hu2021}. Similarly, findings from our work may
  also be used to optimize the training process of graph convolutional
  neural networks.
\end{itemize}

\hypertarget{limitations}{%
\subsection{Limitations}\label{limitations}}

Our analysis is conditioned on the linear low-rank decomposition
assumption of the SVD algorithm. Another matrix factorization algorithm
could have decomposed the truncated \(\truncatelogq\) matrix in such a
way as to learn better node embeddings for link prediction. However, it
would be surprising that the information from only high-frequency pairs
would be sufficient for link prediction. On another note, the original
NetMF \citep{qiu2018} was not tested on link prediction problems, i.e.,
the matrix factorization of the truncated \(\pmi\) (\(\truncatelogq\))
was benchmarked on node classification problems. Lastly, our proposition
of smoothening the information from low-frequency pairs might yield a
highly dense matrix, which might not scale for big graphs.

\hypertarget{conclusion-and-future-work}{%
\subsection{Conclusion and future
work}\label{conclusion-and-future-work}}

Our work contributes to a better understanding of the interplay between
the skip-gram powered neural graph embedding algorithms and matrix
factorization. Neural graph embeddings implicitly decompose matrices
that represent quantities derived from the co-occurrence frequencies in
simulated random walks. We showed that the link prediction accuracy of
graph embeddings strongly depends on the transformations of the original
graph co-occurrence matrix that they decompose. In particular,
smoothening entries corresponding to low-frequency pairs, as opposed to
truncating them, may lead to better performance in link prediction.

Future work will be focused on addressing the limitations and devising
workaround strategies to make sure that our approach can scale to very
big graphs. In particular, we would like to incorporate spectral graph
sparsification \citep{batson2013} into our algorithm.

\hypertarget{acknowledgements}{%
\subsection{Acknowledgements}\label{acknowledgements}}

The computational results presented have been achieved in part using the
Vienna Scientific Cluster (VSC).


\bibliography{biblio}

%

\end{document}